\documentclass[preprint,12pt]{elsarticle}
 \usepackage{graphicx}
 \usepackage{epsfig}
\usepackage{amsmath}
\usepackage{amssymb}

\journal{Computer Physics Communications}

\begin{document}

\begin{frontmatter}

\title{$\Theta$-polymers in crowded media under the stretching force}

\author[itp,icmp]{Viktoria Blavatska}
\ead{blavatska@itp.uni-leipzig.de,viktoria@icmp.lviv.ua}

\address[itp]{Institut f\"ur Theoretische Physik and Centre for Theoretical Sciences (NTZ), Universit\"at Leipzig, Postfach 100\,920,
D-04009 Leipzig, Germany}
\address[icmp]{Institute for Condensed
Matter Physics of the National Academy of Sciences of Ukraine,\\
79011 Lviv, Ukraine}
\author[itp]{Wolfhard Janke\corref{cor1}}
\ead{Wolfhard.Janke@itp.uni-leipzig.de}
\cortext[cor1]{Corresponding author}
\begin{abstract}
We study the peculiarities of stretching of globular polymer macromolecules in a disordered (crowded) environment, using the model of self-attracting self-avoiding
walks on site-diluted percolative lattices in space dimensions $d=3$. 
Applying the pruned-enriched Rosenbluth chain-growth method (PERM),
we construct the phase diagram of collapsed-extended state coexistence when varying temperature and stretching force. The change in shape characteristics of globular polymers under stretching is analyzed as well.

\end{abstract}

\end{frontmatter}

\section{Introduction}

Long flexible polymer macromolecules in a good solvent form crumpled coil configurations which are perfectly captured by the model of self-avoiding random walks (SAW) on a regular lattice  \cite{polymerbook}. This regime holds at temperatures $T$ well above the so-called $\Theta$-point. 
When lowering the temperature, the effect of monomer-monomer attraction grows and the polymer radius shrinks. At $T=T_{\Theta}$, a crossover occurs from high-temperature SAW behaviour to the $\Theta$-statistics. At this particular temperature polymers in $d=3$ dimensions behave effectively as simple random walks (RW).
Below the $\Theta$-temperature, the entropic effects, which make the polymer chain swell, are overcome by interaction energy and a collapse to the globule regime occurs.
The coil-globule transition is considered to be of second order \cite{polymerbook}, in the sense that the density of an infinite globule is zero at $T=T_{\Theta}$ and increases continuously when further lowering the temperature.

The coil-globule transition is of interest in various respects, being deeply connected with problems like protein folding and DNA condensation.
The properties of polymers in the vicinity of the $\Theta$-point
can be successfully studied on the basis of self-attracting
self-avoiding walks (SASAW), where a
nearest-neighbor interaction is included: an attractive energy $\epsilon$ 
 between two neighbor sites is introduced. 
The coil-globule transition of flexible polymers has been so far the subject of
numerous studies \cite{Privman86,Grassberger95,Barkema98,Grassberger97,Brak98,Vogel07}.
Recent numerical estimates give  $T_{\Theta}(d=3)=
3.717(3)$ \cite{Grassberger97}.

In studying the folding dynamics and transport properties of proteins, an important role is played by 
global shape properties of a typical polymer configuration. The asymmetry of polymer shape can be characterized, e.g., by the so-called averaged asphericity $\langle A_d \rangle$ \cite{Aronovitz86,Rudnick86}, 
which takes on a maximum value of one for a completely stretched  configuration,
and equals zero for a spherical form. It was realized experimentally \cite{Dima04,Rawat09} that the majority of globular proteins are characterized by an asphericity value $\langle A_d \rangle \approx 0.1$, thus being almost spherical.

In polymer physics, of great importance is the understanding of the behaviour of macromolecules in the presence of structural disorder. In particular, related problems have been raised recently in studies of  protein folding in the natural cellular environment \cite{Goodesel91,review}. Real biological cells can be described as a very crowded environment built of the biochemical species, which occupy a large fraction of the total volume.
In the language of lattice models, the crowded environment with structural obstacles can be considered as a disordered lattice, where
some amount of randomly chosen sites contains defects.
Of particular interest is the case, when the concentration $p$ of lattice sites allowed for the SAWs
 equals the critical concentration $p_c(d{=}3)=0.31160$ {\cite{Grassberger92}}
 and the lattice becomes percolative.
  It is established that the value of the  $\Theta$-temperature is lowered  due to the presence of disorder
\cite{Barat95,Chakrabarti90,Bhattacharya84,Blavatska09}, numerical estimates give
 $T_{\Theta}^{p_c}(d=3)=0.71(2)$ \cite{Blavatska09}.

The recent progress in experimental techniques  makes it possible to monitor the behaviour of various polymers under tension and stress. In particular, applying a force on
 an isolated protein, the unfolding of the giant titine protein \cite{Rief97} and stretching of collapsed DNA molecules \cite{Baumann00} have been studied.
Of special interest in biophysics is the stretching of  globular polymers below the $\Theta$-point. 
Force not only influences the structural properties of polymers,
 but also may introduce a new completely stretched state which is otherwise not accessible.
 The properties of force-induced  transitions in polymers have been studied intensively  \cite{Halperin91,Goritz96,Grassberger02,Marenduzzo02,Kumar}. The response of a polymer in crowded media to the stretching force modelled by the SASAW model on a percolative lattice has been analyzed recently  in \cite{Kumarpc} and \cite{Blavatska09}. The interesting question about how the shape properties of almost spherical polymer globules are influenced by stretching remains, however, completely unresolved.  

 The aim of the present study is to apply numerical simulations to analyze the properties of SASAWs on site-diluted lattices
at the percolation threshold  under applied external stretching force in space dimensions $d=3$. We analyze the effect of applied force on the phase transitions between collapsed, extended and stretched phases and estimate the influence of stretching on the shape parameters of globular proteins in crowded environments.

\section{The method}

We consider site percolation on regular lattices of edge lengths up to $L_{{\rm max}}{=}200$ in $d=3$. Each site of the lattice was assigned to be occupied with probability $p_c$ and empty otherwise.
To obtain the backbone of a percolation cluster on a given disordered lattice, we apply an algorithm explained in detail in our previous papers
\cite{Blavatska08}. 

To study SASAWs on the backbone of percolation clusters, we apply the pruned-enriched Rosenbluth method (PERM) \cite{Grassberger97},
taking into account that the SASAW can have its steps only on the sites belonging to the backbone of the percolation cluster.  PERM is based on the original Rosenbluth-Rosenbluth (RR) method
\cite{Rosenbluth55} and enrichment strategies \cite{Wall59}. The polymer grows step by step,
i.e., the $n$th monomer is placed at a randomly chosen  empty neighbor site of the last placed $(n-1)$th  monomer ($n\leq N$, where $N$ is the total length of the chain).
The growth is stopped, if the total length of the chain is reached.
A weight
\begin{equation}
W_n=\prod_{l=2}^n m_l {\rm e}^{-\frac{(E_l-E_{l-1})}{k_B T}}
\label{weight1}
\end{equation}
is given to each sample configuration at the $n$th step, where $m_l$ is the number of free lattice sites to place the $l$th monomer, $E_l$ denotes the energy of the $l$-step chain
 ($E_l=z_l\cdot \epsilon$ with  $\epsilon$ being an attractive
energy between two nearest neighbors and $z_l$ the number
 of nearest neighbors contacts for a given chain) and $k_B$ is
 the Boltzmann constant. In what follows, we will assume units in which $\epsilon=-1, k_B=1$.

The configurational averaging for any quantity of interest  then has the form:
\begin{eqnarray}
&&\langle \ldots \rangle=\frac{1}{Z_N}{\sum_{{\rm conf}}W_N^{{\rm conf}}\ldots}, 
\,\,\,\,Z_N=\sum_{{\rm conf}} W_N^{{\rm conf}} \label {R}.
\end{eqnarray}
This method is particularly useful for studying $\Theta$-polymers, since the Rosenbluth weights of the statistically relevant chains approximately cancel
against their  Boltzmann probability. 

Population control in PERM suggests pruning configurations with too small weights, and enriching the sample with copies of high-weight configurations \cite{Grassberger97}. These copies are made while the chain is growing, and continue to grow independently of each other. Pruning and enrichment are performed by choosing thresholds $W_n^{<}$
and $W_n^{>}$ depending on the estimate of the partition sum of the $n$-monomer chain. 
If the current weight $W_n$ of an $n$-monomer chain is less than $W_n^{<}$, a random number $r{=}{0,1}$ is chosen; if $r{=}0$, the chain is discarded, otherwise it is kept and its weight is doubled. Thus, low-weight chains are pruned with probability $1/2$. If $W_n$ exceeds  $W_n^{>}$, the configuration is doubled and the weight of each copy is taken as half the original weight.

In stretching studies, one end of the chain is subjected to an external force
$F$ acting in a chosen direction,
say $x$, while the other end (the starting point) is kept fixed. The stretching energy
 $E_s$ arising due to
the applied force for an  $n$-step trajectory is given by:
$
E_s=- F \cdot x
$,  where $x\equiv x_n-x_0$ denotes the $x$-component of the distance from the
starting point.
The Rosenbluth weight factor $W_n$ is here taken to be:
\begin{equation}
W_n=\prod_{l=2}^n m_l {\rm e}^{-\frac{(E_l-E_{l-1})-F(x_l-x_{l-1})}{k_B T}}.
\label{weight}
\end{equation}

For estimations of quantities of interest we have to perform two types of averaging:  the first over all polymer configurations 
on a single percolation cluster according to  (\ref{R});
the second average is carried out over different realizations of disorder, i.e., over all percolation clusters constructed:
\begin{eqnarray}
&&\overline{\langle \ldots \rangle}{=}\frac{1}{M}\sum_{i{=}1}^M \langle \ldots \rangle_i,\label{av}
\end{eqnarray}
where $M$ is the number of different clusters and the subscript $i$ means that a given quantity is calculated on the cluster $i$.

\section{Results}

The properties of systems in the vicinity of a second-order phase transition can be studied by analyzing the
peak structure of the specific heat $C_V$  as a function of temperature indicating crossovers between physically different states. In the case
of a polymer system, this corresponds to the transition between globule and coil regimes.
$C_V$ can be expressed via energy fluctuations as follows:
 \begin{equation}
C_V(T)=\frac{1}{NT^2}\left(\overline{\langle E^2 \rangle}- \overline{\langle E \rangle^2}\right).
\label{cvperc}
\end{equation}

\begin{figure}[t!]
 \begin{center}
\includegraphics[width=6cm]{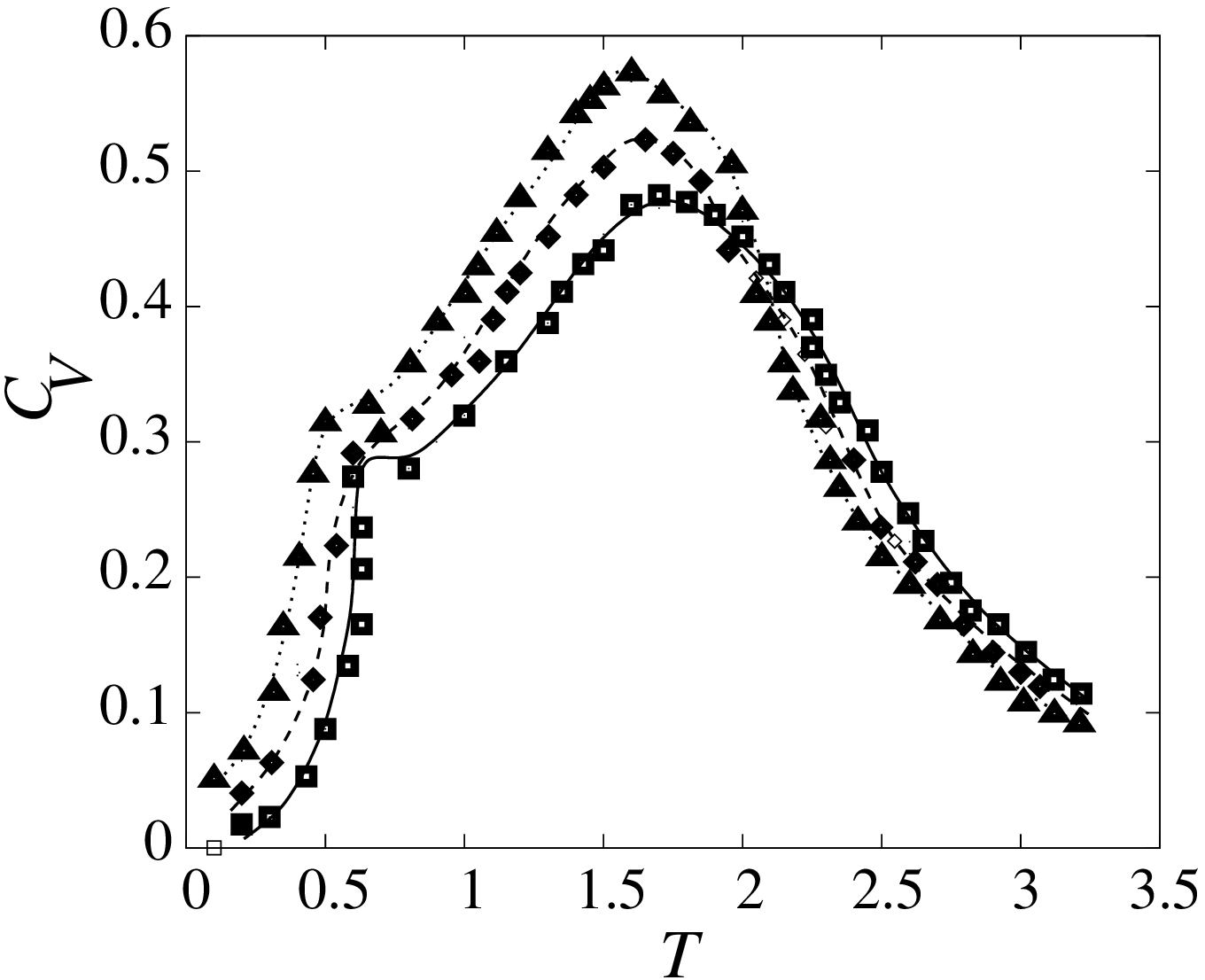}
\includegraphics[width=6cm]{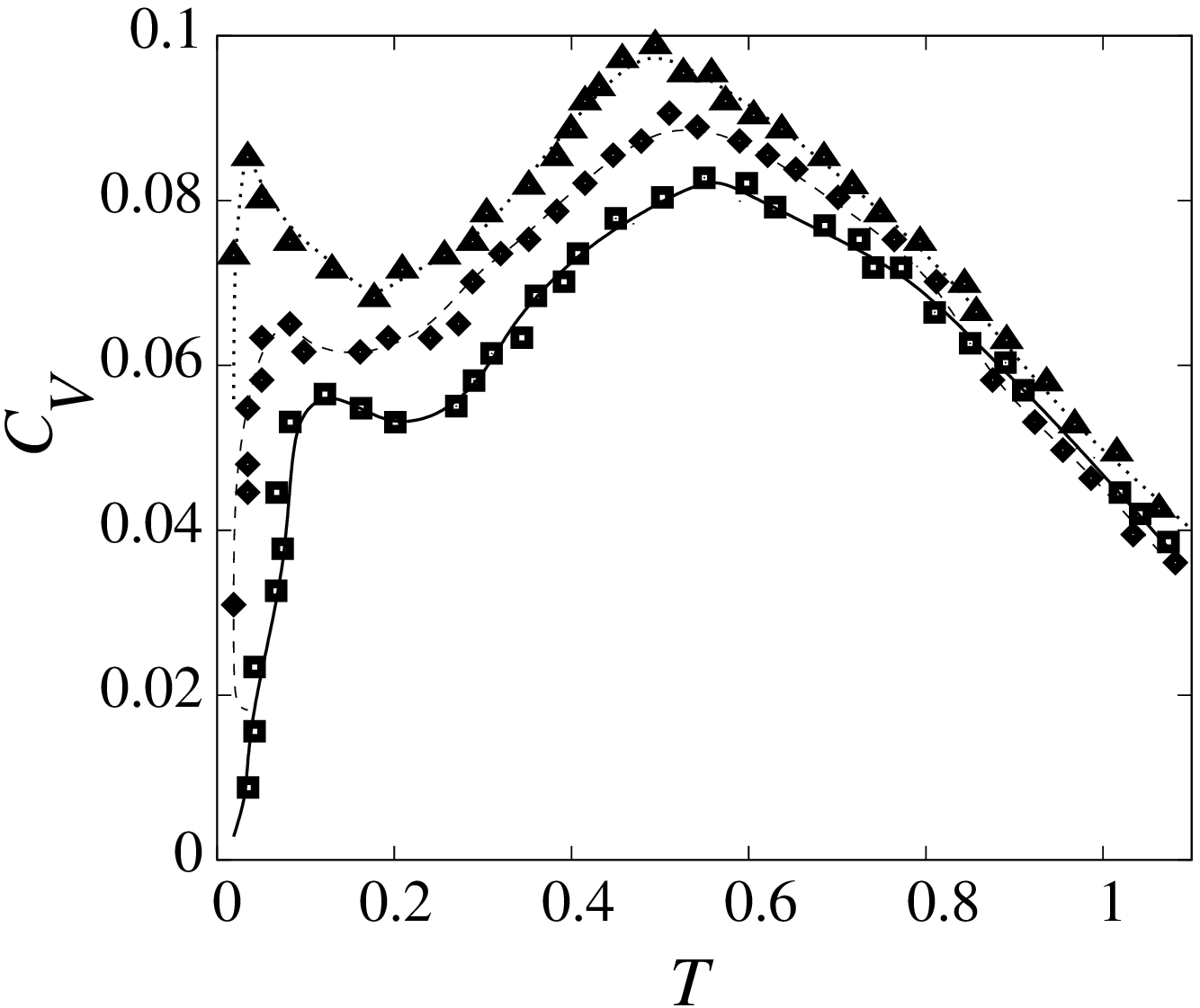}
\end{center}
\caption{\label{cvF3}
Specific heat per monomer of a SASAW with $N=90$ steps in $d=3$ under stretching force $F$ as a function of temperature. Left: pure lattice, right: backbone of percolation cluster. Squares: $F=0.2$, diamonds: $F=0.4$, triangles: $F=0.6$.}
\end{figure}

To study the $\Theta$-transition of SASAWs, when the external stretching force is acting in the environment, we are working in the ``constant-force" ensemble.
Fixing the value of $F$, we study the specific-heat behaviour (Fig. \ref{cvF3}). Analyzing the peak structure of the specific heat, we immediately conclude,
that  increasing the value of $F$ leads to decreasing the transition temperature. With increasing $F$, the averaged energy of the chain decreases, because  applied force stretches the polymer globule. The value of the  transition temperature is thus shifted by the presence of force.

\begin{figure}[t!]
 \begin{center}
\includegraphics[width=6cm]{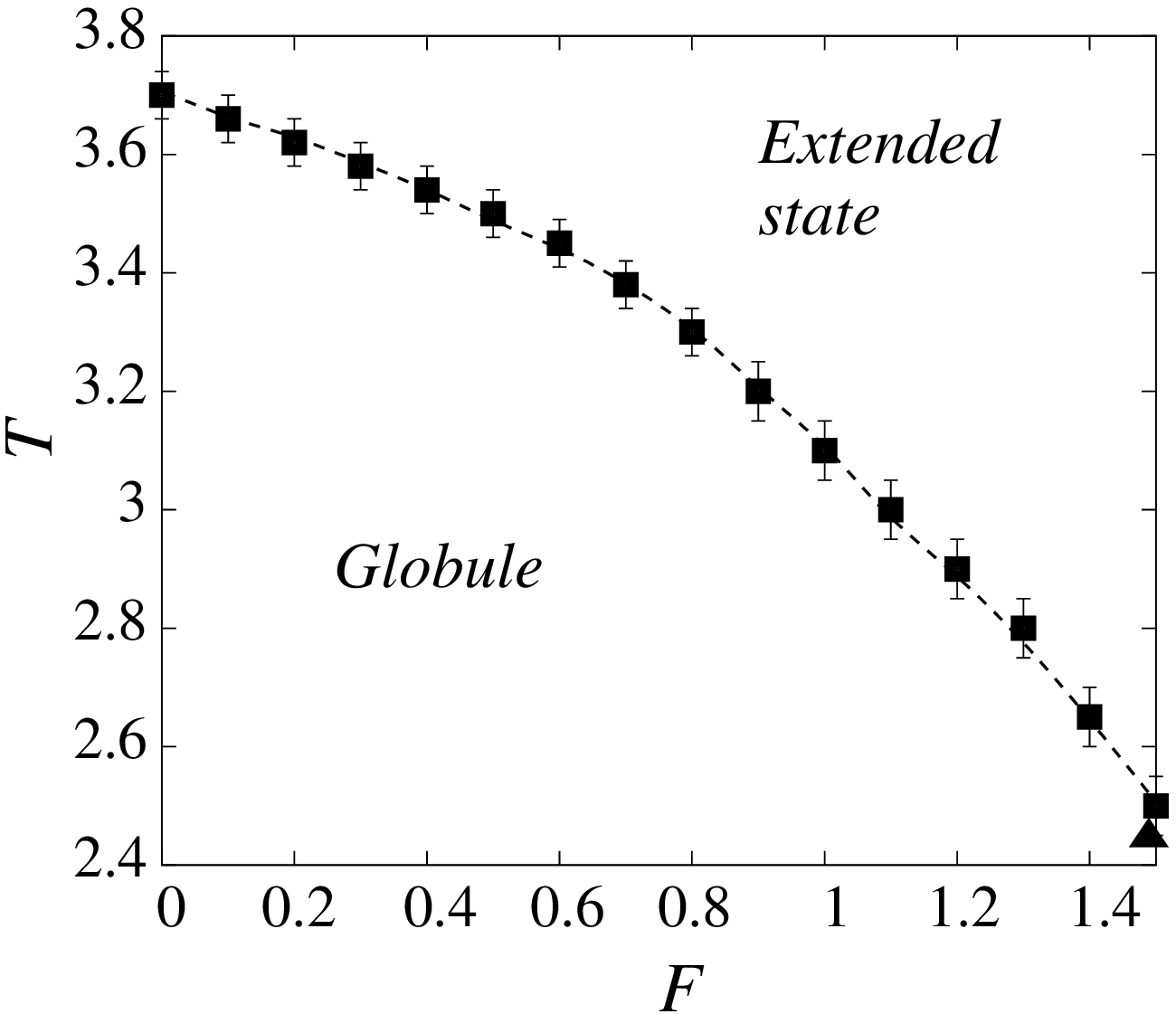}
\includegraphics[width=6cm]{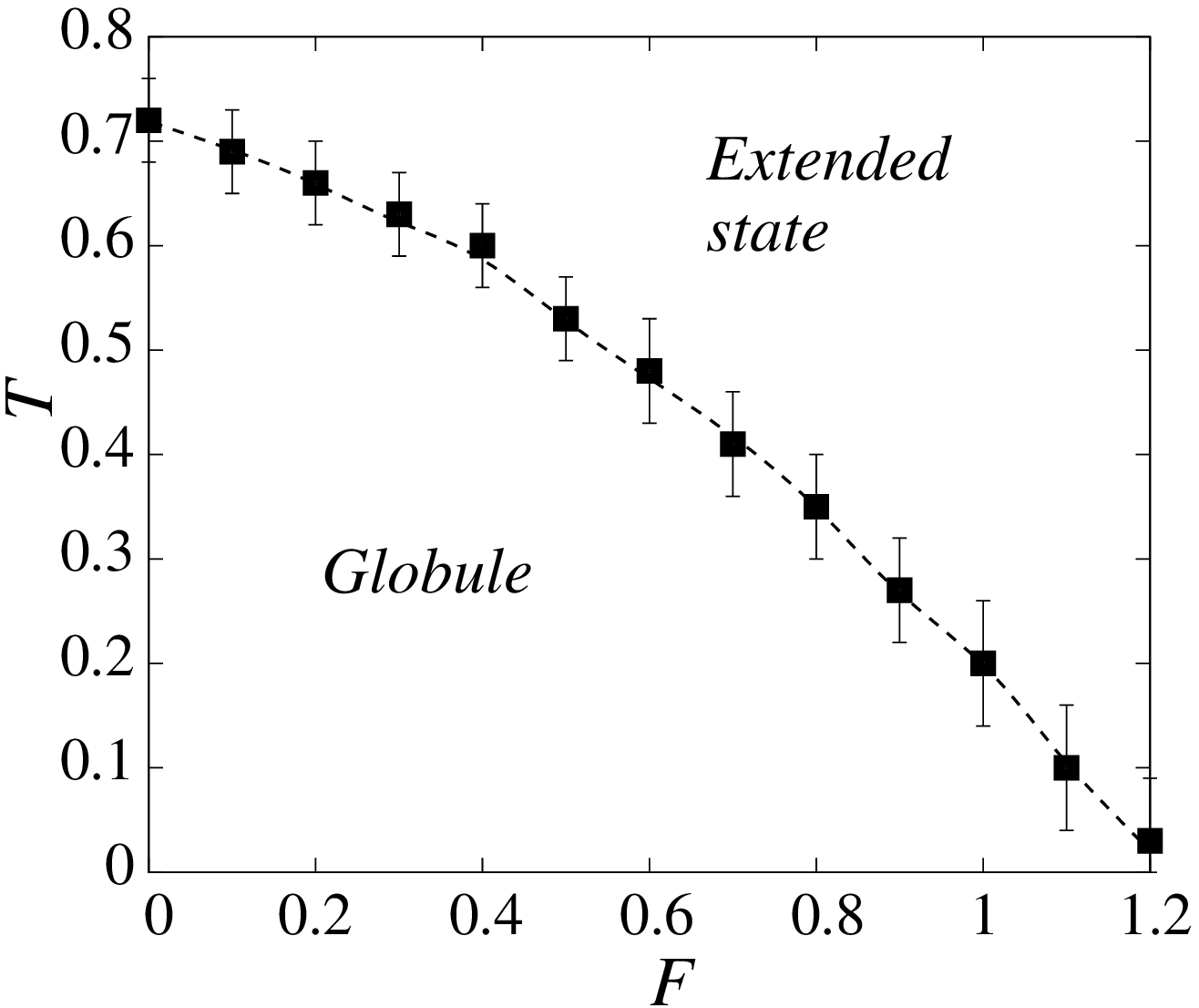}
\end{center}
\caption{\label{cvdiagram} Phase diagrams of a SASAW under applied force $F$ in  $d=3$. Left: pure lattice, right: backbone of percolation cluster. The filled triangle presents the result of Ref. \cite{Grassberger02}, $F=1.5, T_{\Theta}\approx 2.46$. }
 \end{figure}
For finite chain length $N$, the temperature defined by the position of the specific-heat maximum $T_{C_V}^{{\rm max}}(N)$ is well below the collapse transition ${\Theta}$-temperature. This finite-size deviation of $T_{C_V}^{{\rm max}}(N)$ from $T_{\Theta}$ obeys scaling behaviour with $N$:
\begin{equation}
{T_{C_V}^{\rm max}(N)}-{T_{\Theta}}\sim a\cdot N^{-\nu_{\Theta}}+\frac{b}{N},
\label{tcv}
\end{equation}
where $a, b$ are constants and  $\nu_{\Theta}(d=3)=1/2$ is the size exponent of a SAW at the $\Theta$-point.
Our  estimates for $T_{\Theta}^{p_c}$ in the presence of a stretching force are obtained by least-square fitting of (\ref{tcv}) in Ref.  \cite{Blavatska09}.
Results are presented in Fig.~\ref{cvdiagram} in the form of a phase diagram of transitions from globular to extended states.

The measure of the shape properties of a specified spatial conformation of a polymer chain can be characterized \cite{Rudnick86} in terms of the gyration tensor $\bf{Q}$ with components: 
$$
Q_{ij}=\frac{1}{N}\sum_{n=1}^N(x_n^i-{x^i_{CM}})(x_n^j-{x^j_{CM}}),\,\,\,\,i,j=1,\ldots,d,
$$
where $x_n^{j}$ is the $j$th coordinate of the position vector of the $n$th monomer of a polymer chain ($n=1,\ldots,N$), and ${x^j_{CM}}=\sum_{n=1}^Nx_n^j/N$ is the coordinate of the center-of-mass position vector. 
The extent of asphericity of a polymer configuration can be characterized by the quantity ${A_d}$ defined as \cite{Aronovitz86}:
\begin{equation}
{A_d} =\frac{1}{d(d-1)} \sum_{i=1}^d\frac{(\lambda_{i}-{\overline{\lambda}})^2}{\overline{\lambda}^2}=
\frac{d}{d-1}\frac{\rm {Tr}\,\bf{\hat{Q}}^2}{(\rm{Tr}\,{\bf{Q}})^2}, \label{add}
\end{equation}
with $\lambda_i$ being the eigenvalues of the gyration tensor, ${\overline{\lambda}}\equiv {\rm Tr}\, {\bf{Q}}/d$, 
and ${\bf{{\hat{Q}}}}\equiv{\bf{Q}}-\overline{\lambda}\,{\bf{I}}$ (here $\bf{I}$ is  the unity matrix).
This universal quantity equals zero for a spherical configuration, where all the eigenvalues are equal, $\lambda_i=\overline{\lambda}$, and takes a maximum value 
of one in the case of a rod-like configuration, where all the eigenvalues equal zero except of one. Thus, the inequality $0\leq A_d\leq 1$ holds.

\begin{figure}[t!]
 \begin{center}
\includegraphics[width=6.2cm]{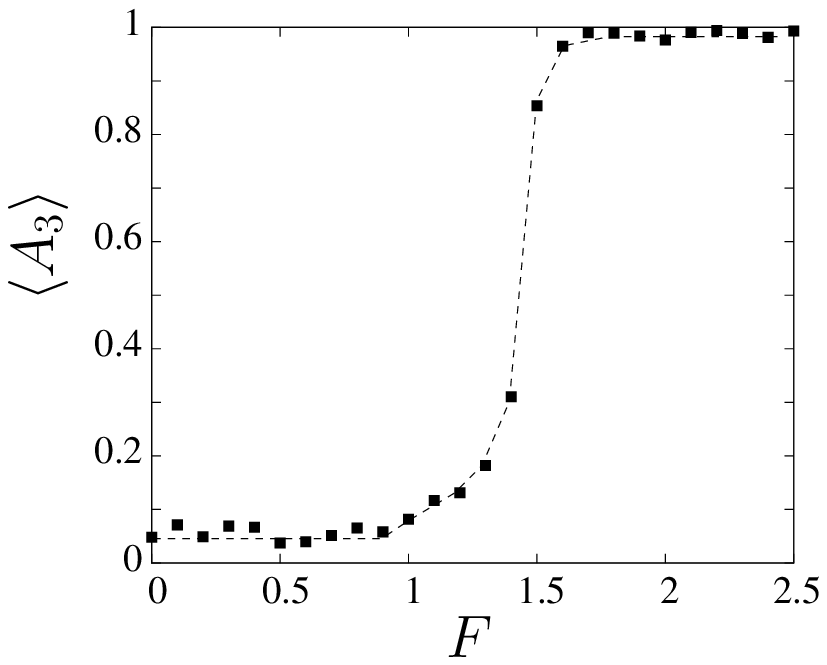}
\includegraphics[width=6.5cm]{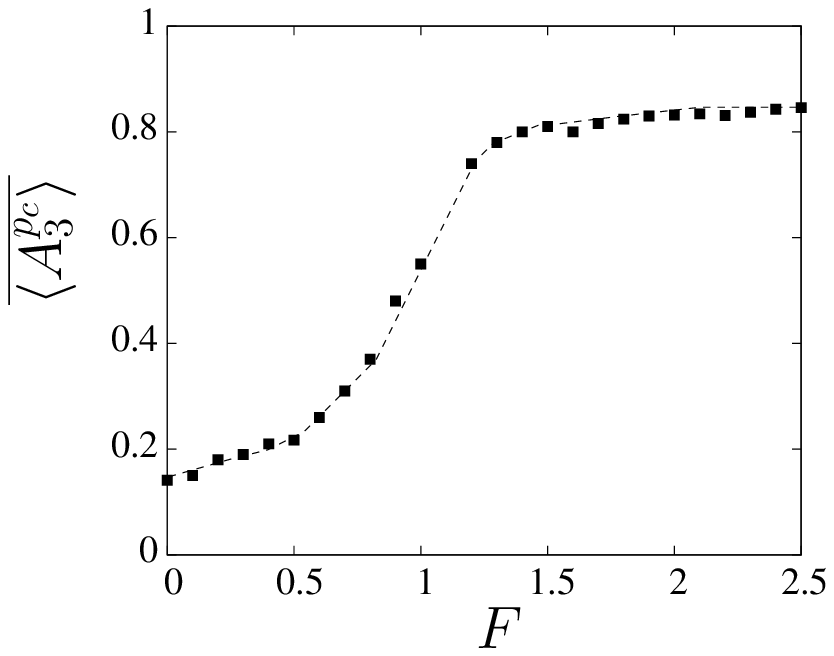}
\end{center}
\caption{\label{Ad} Averaged asphericity of an $N=90$-step SASAW as function of applied force $F$ in  $d=3$. Left: pure lattice ($T=1.8$), right: backbone of percolation cluster ($T=0.2$). Lines are guides to the eyes.  }
 \end{figure}

In Fig. \ref{Ad} we present the averaged asphericity of SASAWs, giving information about the internal structure of the polymer configuration under applied force,  at a temperature well below the $\Theta$-point for the cases of a pure lattice and the backbone of a percolation cluster.
Note that in the absence of force, $\langle A_3 \rangle$ for SAWs on a pure lattice is very close to zero, whereas in the disordered case, due to the complicated structure of the underlying  percolative lattice, globular configurations are more elongated with larger ${\overline {\langle A_3^{p_c}\rangle}}$ values. 
At small $F$, a polymer chain is still in the compact folded state and is just slightly oriented along the force direction. Under inreasing forces, the polymer chain takes on  a conformation similar to the extended (swollen) structure. Note, that completely stretched states, corresponding to $\langle A_d \rangle \simeq 1$ can be obtained only in the pure case and are not accessible on the percolative lattices due to the complicated fractal structure on the underlying percolation cluster.

\section{Conclusions}

We studied  self-attracting self-avoiding walks on disordered lattices
in space dimensions $d=3$, modelling flexible polymer macromolecules in porous environment.
We considered the special case, when the concentration of disorder is
exactly at the percolation threshold, so that an incipient percolation cluster of sites, allowed for SAWs, emerges on the lattice.
Keeping one end of a SASAW trajectory on the backbone of a percolation cluster fixed,
we applied a stretching force $F$, acting in some chosen direction (say, $x$). Based on our numerical simulation data, we constructed phase diagrams of collapsed and extended states coexistence. 
The behaviour of averaged asphericity of globular SASAWs on a percolation cluster under a stretching force was analyzed as well. As expected, in the case of a pure lattice, stretching leads to increasing the asymmetry indicating at some critical value of $F$ a transition to a completely stretched state with $\langle A_d \rangle \approx 1$. For SASAWs on percolative lattices, the completely stretched states are not accessible due to the complicated fractal structure of the underlying percolation cluster even under very strong stretching. 

\section*{Acknowledgement}
V.B. is grateful for support through the S\"achsische DFG-Forschergruppe FOR877.

\end{document}